\documentstyle[12pt,psfig,aasms4]{article}
\def\paren#1{\left( #1 \right)}

\newcommand{\gtsima}{$\; \buildrel > \over \sim \;$}
\newcommand{\ltsima}{$\; \buildrel < \over \sim \;$}
\newcommand{\simgt}{\lower.5ex\hbox{\gtsima}}
\newcommand{\simlt}{\lower.5ex\hbox{\ltsima}}

\newcommand{\A}{ {\scriptscriptstyle {\rm A}} }
\newcommand{\B}{ {\scriptscriptstyle {\rm B}} }
\newcommand{\X}{ {\scriptscriptstyle {\rm X}} }
\def\pp{\par\parshape 2 0truecm 15.5truecm 1truecm 14.5truecm\noindent}
\begin{document}

\baselineskip=20pt

\vspace*{-1.5cm}
\begin{minipage}[c]{3cm}
  \psfig{figure=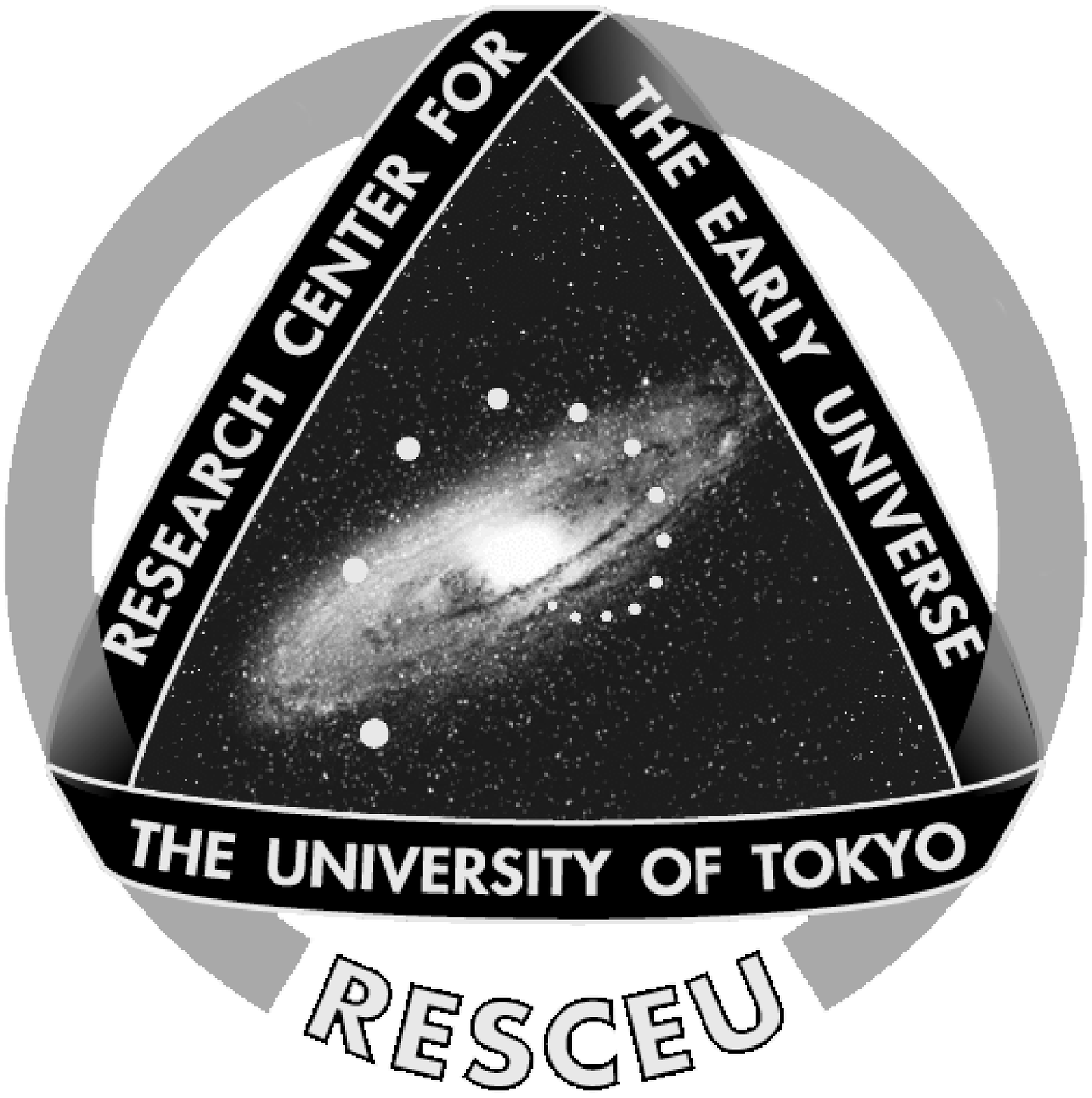,height=3cm}
\end{minipage}
\begin{minipage}[c]{9cm}
\begin{centering}
{
\vskip 0.1in
{\large \sf 
THE UNIVERSITY OF TOKYO\\
\vskip 0.1in
Research Center for the Early Universe}\\
}
\end{centering}
\end{minipage}
\begin{minipage}[c]{3cm}
\vspace{2.5cm}
RESCEU-25/96\\
UTAP-242/96
\end{minipage}\\
\vspace{1.0cm}


\baselineskip20pt
\parskip 2pt

\centerline{\large \bf Re-examination of the Hubble Constant}
\centerline{\large \bf from the Sunyaev-Zel'dovich Effect:}
\centerline{\large \bf Implication for Cosmological Parameters}

\bigskip

\centerline{Shiho Kobayashi$^{1,2}$, Shin Sasaki$^{3}$ and 
Yasushi Suto$^{1,4}$}

\begin{center}
$^{1}$ Department of Physics, The University of Tokyo,
Tokyo 113, Japan\\
$^{2}$ present address:
The Racah Institute for Physics,\\
The Hebrew University, Jerusalem 91904, Israel\\
$^{3}$ Department of Physics, Tokyo Metropolitan University, Hachioji,
Tokyo 192-03, Japan\\
$^{4}$RESCEU (Research Center for the Early Universe),
School of Science, \\  The University of Tokyo, Tokyo 113, Japan\\
\end{center}

\centerline{e-mail: shiho@alf.fiz.huji.ac.il,
sasaki@phys.metro-u.ac.jp, suto@phys.s.u-tokyo.ac.jp
}

\received{1996 August 22}
\accepted{1996 September 26}

\bigskip
\centerline{\bf Abstract}

  We performed a systematic and comprehensive estimate of the Hubble
  constant $H_0$ via the Sunyaev-Zel'dovich temperature decrement and
  the X-ray measurements for clusters up to the redshift $z=0.541$,
  with particular attention to its dependence on the density parameter
  $\Omega_0$, and the cosmological constant $\lambda_0$.  The
  resulting values of $H_0$ are largely consistent with that derived
  from the Cepheid distance for nearby galaxies.  We discuss how the
  angular diameter distance $d_\A$ vs. $z$ diagram from the accurate
  SZ measurements of high-$z$ clusters can distinguish the values of
  $\Omega_0$ and $\lambda_0$.

\bigskip

\vfill
\centerline{\it Publications of the Astronomical Society of Japan
  (Letters), vol.48, No.6, in press.}
\newpage

\baselineskip 20pt
\section{Introduction}

There have been a lot of attempts to determine the Hubble constant
$H_0$ ($\equiv 100h\;{\rm km \, s^{-1}\, Mpc^{-1}}$) from the
Sunyaev-Zel'dovich (SZ) temperature decrement and the X-ray
measurements (Sunyaev \& Zel'dovich 1972; see Rephaeli 1995 for a
review). The current results are subject to several uncertainties due
to simplified model assumptions in describing the realistic cluster
gas distribution as well as the limited spatial resolution of the
X-ray and radio observations (Birkinshaw, Hughes \& Arnaud 1991;
Inagaki, Suginohara \& Suto 1995).  Nevertheless this method is very
important for two reasons; it does not rely on any local (or
empirical) calibrator, and it can be used even at high redshift $z
\simgt 1$ in principle where no other reliable distance indicator is
known, perhaps except for the SNe Ia (Perlmutter et al. 1996).

As discussed by Silk \& White (1978) and Birkinshaw et al. (1991), the
observation of the SZ effect determines the angular diameter distance
$d_\A (z)$ to the redshift $z$ of the cluster.  For $z \ll 1$,
$d_\A(z)$ is basically given only in terms of $H_0$ ($\sim cz/H_0$,
with $c$ being the light velocity).  If $z \simgt 0.1$, however, the
density parameter $\Omega_0$, the dimensionless cosmological constant
$\lambda_0$, and possibly the degree of the inhomogeneities in the
light path (see \S 3 below; Dyer \& Roeder 1972) make
significant contribution to the value of $d_A(z)$ as well.

In addition, the current estimate of $H_0$ from optical observations
is converging with fractional uncertainty of about 10 percent level,
$h=0.7\pm0.1$ (Freedman et al. 1994; Tanvir et al. 1995).  This
implies that one may place useful limits on $\Omega_0$ and $\lambda_0$
if accurate and reliable SZ observations are carried out {\it even for
  a single high-$z$ cluster}.  Unfortunately the non-spherical
distribution and clumpiness of the cluster gas among others would
inevitably contaminate any SZ observation (Birkinshaw et al. 1991;
Inagaki et al.  1995), and this is why we have to proceed
statistically.  In this {\it Letter}, we present estimates of $H_0$
for seven SZ clusters (Table 1) by taking proper account of the effect
of $\Omega_0$ and $\lambda_0$ for the first time.

\section{Angular Diameter Distance from the SZ Effect}

The temperature decrement in the cosmic microwave background
(CMB) by the SZ effect observed at an angle $\theta$ from the center
of a cluster is given by the integral over the line of sight:
\begin{eqnarray}
  \frac{\Delta T}{T_0}(\theta)&=& -2\xi(x)
    \int^{\infty}_{-\infty}\frac{k_B T_{gas}}{m_e c^2}\sigma_T n_e dl ,
\label{eq:dttsz} \\
\xi(x) &=&
\frac{x^2e^x}{2(e^x-1)^2}\paren{ 4- x\coth\frac{x}{2}},
\end{eqnarray}
where $T_{gas}$ and $n_e$ are the temperature and the number density
of the electron gas in the cluster, $T_0$ is the temperature of the
CMB ($= 2.726$K) and $x$ is the dimensionless
frequency $h_p \widetilde\nu / k_B T_0$ ($\widetilde\nu$ is the
frequency of the radio observation, $h_p$ is the Planck constant), and
the other symbols have their usual meanings. The relativistic
correction to the above formula is negligible for $T_{\rm gas} \simlt
15$keV and $x \simlt 1$ (e.g., Rephaeli 1995).

Similarly the X-ray surface brightness $S_\X$ of the cluster observed
at $\theta$ in the X-ray detector bandwidth $\nu_1 \sim \nu_2$ is
written as
\begin{equation}
  S_\X(\theta)=\frac{1}{4\pi(1+z)^4}
\int^{\infty}_{-\infty} dl \, \int^{\nu_2(1+z)}_{\nu_1(1+z)}
                               \frac{d^2 L_\X}{d\nu dV}d\nu ,
\label{eq:sx}
\end{equation}
where $z$ is the redshift of the cluster, and $d^2L_\X/d\nu dV$ is the
X-ray emissivity per unit frequency. If $T_{gas}$ is greater than
several keV, the line contribution is negligible\footnote{ The X-ray
  flux due to line emission for a solar abundance plasma at $T_{\rm
    gas}= 6.8$keV, for example, is about 6 percent of that of
  bremsstrahlung in the ROSAT band. } and $S_\X$ is well approximated
by that of the thermal bremsstrahlung:
\begin{eqnarray}
\frac{d^2L_\X}{d\nu dV} &=&
\alpha(T_{gas}) \, n^2_e \,
\overline{g}_{ff}(T_{gas},\nu)
\exp\left(-\frac{h_p\nu}{k_\B T_{gas}}\right) ,
\label{eq:xemissivity} \\
\alpha(T_{gas}) &\equiv& 
\frac{2^5\pi e^6}{3m_e c^2}\paren{\frac{2\pi}{3m_e c^2 k_B T_{gas}}}^{1/2}
\frac{2}{1+X} .
\label{eq:alpha}
\end{eqnarray}
We assume the primordial abundances of hydrogen and helium with the
hydrogen mass fraction $X=0.755$. The velocity averaged Gaunt factor
of the thermal bremsstrahlung, $\overline{g}_{ff}(T_{gas},\nu)$, is
computed according to Kellogg, Baldwin \& Koch (1975).

The conventional assumption is that the cluster gas follows a
spherically symmetric isothermal $\beta$-model:
$n_e(r)=n_{e0}/(1+r^2/d^2_\A(z)\theta^2_c)^{3\beta/2}$, where
$\theta_c$ and $n_{e0}$ are the angular core radius (fitted from the
radio or X-ray imaging) and the central density of the cluster gas,
and $d_\A(z)$ denotes the angular diameter distance to the redshift
$z$ of the cluster. Then one can integrate equations (\ref{eq:dttsz})
and (\ref{eq:sx}) over the line of sight, and their central values
($\theta=0$) reduce to
\begin{eqnarray}
&&\frac{\Delta T(0)}{T_0} = 
-2\xi(\nu)
\sqrt{\pi}\sigma_T
\frac{k_\B T_{gas}}{m_ec^2}
\frac{\Gamma(3\beta_r/2-1/2)}{\Gamma(3\beta_r/2)}n_{e0}d_\A(z)\theta_{r} ,
\label{eq:dec0}\\
&&S_\X(0) = 
\frac{\alpha(T_{gas})k_\B T_{gas}}{h_p}
\frac{n_{e0}^2 d_\A(z)\theta_\X}{4\sqrt{\pi}(1+z)^4}
\frac{\Gamma(3\beta_\X-1/2)}{\Gamma(3\beta_\X)}
\int^{h_p\nu_2(1+z)/k_\B T_{gas}}_{h_p\nu_1(1+z)/k_\B T_{gas}}
\overline{g}_{ff}{\rm e}^{-x} dx ,
\label{eq:sx0}
\end{eqnarray}
where $\Gamma(x)$ denotes the gamma function.  Although the values of
$\beta$ and $\theta_c$ obtained by the radio ($\beta_r$ and
$\theta_r$) and the X-ray observations ($\beta_\X$ and $\theta_\X$)
should be identical as long as the spherical $\beta$-model
is exact, different values are often reported in the
literature.  While we distinguish them in the above equations
following the literature, but the difference due to the different
choices of the parameters should be regarded as an additional
uncertainty in the present context (see \S 3 below).

Finally equations (\ref{eq:dec0}) and (\ref{eq:sx0}) can be solved for
$d_\A$:
\begin{eqnarray}
d_\A &=&
\frac{\alpha(T_{gas}) k_B T_{gas}}{16\pi^{3/2}h_p\sigma_T^2(1+z)^4}
\paren{\frac{m_e c^2}{k_BT_{gas}}}^2
\frac{\Gamma(3\beta_\X-1/2)}{\Gamma(3\beta_\X)}
\paren{\frac{\Gamma(3\beta_r/2)}{\Gamma(3\beta_r/2-1/2)}}^2 \nonumber \\
\qquad  &\quad& \times
\frac{(\Delta T/T(0))^2}{\xi^2(\nu)S_\X(0)}
\frac{\theta_\X}{\theta_r^2}
\int^{h_p\nu_2(1+z)/k_\B T_{gas}}_{h_p\nu_1(1+z)/k_\B T_{gas}}
\overline{g}_{ff} {\rm e}^{-x} dx .
\label{eq:obsda}
\end{eqnarray}
Comparison of the above estimate with the theoretical formula for
$d_\A$ in the Friedman-Robertson-Walker model, yields a relation among
$H_0$, $\Omega_0$ and $\lambda_0$ (e.g., Silk \& White 1978; Inagaki
et al. 1995) as we discuss extensively below.

\section{$H_0$, $\Omega_0$ AND $\lambda_0$ from a Sample of SZ Clusters}

In order to estimate $d_\A$ and see the constraints on $H_0$,
$\Omega_0$ and $\lambda_0$, we use seven clusters of which we can
collect the necessary data in radio and X-rays for the analysis from
the literature and the private communications\footnote{We did not
  include A665 (Birkinshaw et al. 1991) since we could not find the
  proper X-ray data from the literature. }.  In the case of CL0016+16,
we adopt the value of $S_\X(0)$ given in White, Silk \& Henry (1981)
and Birkinshaw, Gull, \& Moffet (1981). Otherwise we first compute
$S_\X(0)$ in equation (\ref{eq:sx0}) from the value of $n_{e0}$ given
in the literature of the X-ray observation, and then estimate $d_\A$
(eq.[\ref{eq:obsda}]) so as not to include the line emission
contribution (even though it is small).  Our estimates of $H_0$ listed
in Table 1 are for the $\Omega_0=1$ and $\lambda_0=0$ model, just for
example; they are in good agreement with the previous work as long as
we adopt the same parameters. In what follows, we assign $1\sigma$
errors simply by taking account of those of $n_{e0}$, $T_{gas}$ and
$\Delta T(0)$ in quadrature. The errors are dominated by the
uncertainty in the SZ temperature decrement.  In reality those values
would be strongly dependent on the fitted parameters $\beta$ and
$\theta_c$, and thus it is likely that our quoted error bars
underestimate the actual ones.

Figure \ref{fig:h0omega0} illustrates the dependence of the estimated
$H_0$ on $\Omega_0$ for the two highest-$z$ clusters in Table 1. More
specifically, for CL0016+16 ($z=0.541$) and A2163 ($z=0.201$), $h=1.03
^{+0.59}_{-0.28}$ and $h=0.82^{+0.56}_{-0.24}$ in the $\Omega_0=1$ and
$\lambda_0=0$ model, and $h=1.43^{+0.82}_{-0.38}$ and
$h=0.94^{+0.64}_{-0.27}$ in the $\Omega_0=0$ and $\lambda_0=1$ model,
respectively. Figure \ref{fig:h0z} summarizes the estimated $H_0$ for
all clusters in models with four different sets of $\Omega_0$ and
$\lambda_0$.  The derived values except for A478 are consistent with
$h=0.7\pm0.1$ from the HST Cepheid distance (Tanvir et al. 1995). Note
that if one adopts incorrect cosmological models, the result would
show an {\it artificial} $z$-dependence of $H_0$ (Suto, Suginohara \&
Inagaki 1995).  Since CL0016+16 is at the highest $z$ in our sample,
we plot three different estimates of $H_0$ using different sets of
$\beta$ and $\theta_c$ from radio and X-ray observations (in open
pentagon, open triangle and filled triangle).  The results are quite
sensitive to the choice indicating that the higher resolution image
analysis both in radio and X-ray observations is highly desired (see
aslo Neumann\& B\"{o}hringer 1996)\footnote{In fact, a recent detailed
  analysis using the ROSAT image for CL0016+16, especially taking
  account of the ellipsoidal shape and its temperature structure,
  indicated a substantially smaller value for $H_0$ than we derived
  here (Birkinshaw 1996).}.

Finally Figure \ref{fig:daz} compares the angular diameter
distances $d_\A$ estimated from the SZ effect with several theoretical
ones. Short dash -- long dash line plots the approximation of $d_\A =
cz H_0^{-1}(1-3z/2)$ valid only for $z \ll 1$ and $q_0=0$, simply to
illustrate how it deviates from the correct one; clearly at $z\simgt
0.1$ one should treat $\Omega_0$ and $\lambda_0$ differently and the
constraints cannot be given only in terms of $q_0$.  Also plotted
(thin solid line) is the Dyer-Roeder (or empty beam) distance in
the $\Omega_0=1$ and $\lambda_0=0$ model (Dyer \& Roeder 1972)
so as to indicate the possible effect of the inhomogeneity in the
light propagation (compare with the filled beam case in thick solid
line); in the light of the current systematic and statistical errors
of the SZ measurement, the latter effect is safely neglected.

\section{Discussion and Conclusions}

We have estimated the Hubble constant $H_0$ from seven SZ clusters by
taking proper account of the effect of the cosmological parameters,
which has been neglected previously.  The estimated $H_0$ varies
around 30 percent depending on the assumed $\Omega_0$ and $\lambda_0$
for the two highest-$z$ clusters.  Considering the uncertainties of
the SZ effect (Birkinshaw et al. 1991; Inagaki et al.  1995) and the
observed scatters in the data, however, it is definitely premature to
draw any strong conclusion at this point. Nevertheless it is
reassuring that our estimates are consistent with Cepheid-based
estimate $h=0.7\pm0.1$ (Tanvir et al. 1995) within $2\sigma$
error-bars. In addition, the $d_\A$ vs. $z$ diagram from the future SZ
measurements may provide a promising insight into the values of
cosmological parameters as the Hubble diagram of distant SNe Ia
(Perlmutter et al. 1996).

We thank Mark Birkinshaw for various useful comments on the earlier
manuscript.  We are also grareful to Steve Allen, Akihiro Furuzawa,
Mike Jones, Ian McHardy and Steven T. Myers for correspondences on the
X-ray and radio measurements of the clusters that we analysed.  This
research was supported in part by the Grants-in-Aid by the Ministry of
Education, Science, Sports and Culture of Japan (07CE2002) to RESCEU
(Research Center for the Early Universe).

\newpage

\baselineskip=20pt
\parskip2pt
\bigskip
\centerline{\bf References}
\bigskip

\def\apjpap#1;#2;#3;#4; {\pp#1, {#2}, {#3}, #4}
\def\apjbook#1;#2;#3;#4; {\pp#1, {#2} (#3: #4)}
\def\apjppt#1;#2; {\pp#1, #2.}
\def\apjproc#1;#2;#3;#4;#5;#6; {\pp#1, {#2} #3, (#4: #5), #6}

\apjpap Arnaud, M. et al, 1992;ApJ;251;L69;
\apjppt Birkinshaw, M. 1996; private communication;
\apjpap Birkinshaw, M., Gull, S.F. \& Moffet, A.T. 1981;
ApJ;251;L69;
\apjpap Birkinshaw, M., Hughes, J.P. \& Arnaud, K.A. 1991;
ApJ;379;466;
\apjpap Birkinshaw, M. \& Hughes, J.P. 1994;
ApJ;420;33;
\apjpap Carlstrom, J.E., Joy, M. \& Grego, L. 1996;
ApJ;456;L75;
\apjpap Dyer, C.C. \& Roeder, R.C. 1972;ApJ;174;L115;
\apjpap Elbaz, D., Arnaud, M. \& B\"{o}hringer, H. 1995;
Astro. Astrophys.;293;337;
\apjpap Freedman, W.L. et al. 1994;Nature;371;757 ;
\apjpap Inagaki, Y., Suginohara, T. \& Suto, Y. 1995; PASJ;47;411;
\apjpap Kellogg E., Baldwin J.R. \& Koch, D. 1975; ApJ; 199;299;
\apjppt Myers, S.T., Baker, J.E., Readhead, A.C.S., 
Leitch, E.M. \& Herbig, T. 1996; ApJ, submitted;
\apjppt Neumann, D.M.  \& B\"{o}hringer, H. 1996;MNRAS, in press;
\apjppt Perlmutter, S. et al. 1996; ApJ, in press;
\apjpap Rephaeli, Y. 1995;ARA\&A;33;541;
\apjpap Silk, J. \& White, S.D.M. 1978; ApJ;226;L103;
\apjpap Sunyaev, R.A. \&  Zel'dovich,
Ya.B. 1972;Commts. Astrophys. Space Phys.;4;173;
\apjpap Suto,Y., Suginohara, T. \& Inagaki, Y. 1995;
Prog. Theor. Phys.;93;839;
\apjpap Tanvir, N.R., Shanks, T., Ferguson, H.C. \& Robinson,
D.R.T. 1995; Nature; 377; 27;
\apjpap White, S.D.M., Silk, J. \& Henry, J.P. 1981; ApJ;251;L65;
\apjpap Wilbanks, T.M., Ade, P.A.R., Fischer, M.L.,
Holzapfel, W.L. \& Lange, A.E. 1994;
ApJ;427;L75;
\apjppt Yamashita, K. 1994; in {\it New Horizon of X-ray Astronomy --
First Results from ASCA}, edited by F. Makino and T. Ohashi, (Universal
Academy Press, Tokyo, 1994), 279;

{\small
\noindent
\begin{table}
\caption{Data for the SZ clusters}
\begin{center}
\hspace*{-1cm}
\begin{tabular}{|l|lllllll|}
\hline
 & $T_{gas}$ & -$\Delta T(0)$ & $\beta_\X$ & $\theta_\X$
&$n_{e0}$& $H_0$& ref. \\
\hline
 &  keV &  mK  &  & arcmin  &$ 10^{-3}h^{1/2}{\rm cm}^{-3}$& km/s/Mpc&\\
\hline \hline
CL 0016&$8.4\pm1.2$    &$0.687\pm0.062$
       &2/3            &0.7
       & e)
       &$103^{+59}_{-28}$      & * \\
$z=0.541$&$8.4^{+1.2}_{-0.6}$& $1.584\pm0.256$
       &0.5                &0.5
       & e)
       &$41^{+15}_{-12}$    & 1 \\
       &               &              
       &0.5            &$0.5\pm0.3$
       & e)
       &               & 2\\
       &               &$0.687\pm0.062^{a}$
       &2/3            &0.7
       &
       &               & 3\\
\hline
A2163  
       &$14.6\pm0.5$    & $0.735\pm0.144$
       &$0.62$          &$1.2$    
       &$10.9\pm0.4^{f}$
       & $82^{+56}_{-24}$      & *\\
$z=0.201$&$14.6\pm0.55$   &
       &$0.62\pm0.01$   &$1.20\pm0.05$
       &
       &$82^{+35}_{-22}$  & 4\\
       &$14.6\pm0.5$&    
       &$0.62^{+0.009}_{-0.012}$&$1.2\pm0.046$
       &$10.9\pm0.4^{f}$
       &                 & 5\\
       &            &$0.735^{+0.144}_{-0.126}{}^{\ \ b}$
       &$0.59^{+0.07}_{-0.05}$ &$1.15^{+0.33}_{-0.27}$
       &
       &                 & 6\\
\hline
A2218  
       &$6.7\pm0.45$ &$0.62\pm0.08$
       &$0.65$      &$1.0$
       &$8.9$
       &$60^{+24}_{-13}$   & * \\
$z=0.171$&$6.7\pm0.45$    &$0.62\pm0.08^{c}$
       &$0.65^{+0.05}_{-0.03}$&$1.0\pm0.2$
       &$8.9$
       &$65\pm25$       & 7\\
\hline
A2142  
       &$8.68\pm0.12$    &$0.878\pm0.050$
       &$1.0$            &$3.69$
       &$6.97\pm0.41$
       &$51^{+10}_{-7}$      & *\\
$z=0.0899$&$8.68\pm0.12$    &$0.878\pm0.050^{d}$
       &$1.0\pm0.3$      &$3.69\pm0.14$
       &$6.97\pm0.41$
       &$48^{+43}_{-29}$ & 8\\
\hline
A478   
       &$6.56\pm0.09$    &$0.906\pm0.068$
       &$0.667$          &$1.93$&
       $9.55\pm1.75$
       &$33^{+22}_{-9}$      & *\\
$z=0.0881$&$6.56\pm0.09$    &$0.906\pm0.068^{d}$
       &$0.667\pm0.029$  &$1.93\pm0.30$
       &$9.55\pm1.75$
       &$32^{+18}_{-14}$ & 8\\
\hline

A2256  
       &$7.51\pm0.11$ &$0.397\pm0.047$
       &$0.795      $ &$5.33$ 
       &$3.55\pm0.18$
       &$74^{+26}_{-15}$   & *\\
$z=0.0581$&$7.51\pm0.11$ &$0.397\pm0.047^{d}$
       &$0.795\pm0.020$ &$5.33\pm0.20$
       &$3.55\pm0.18$
       &$72^{+22}_{-19}$ & 8\\
\hline
Coma 
     &$9.10\pm0.40$    &$0.505\pm0.092$
     &$0.75$           &$10.5$
     &$4.09\pm0.06$ 
     &$63^{+28}_{-15}$      & *\\
$z=0.0235$& $9.10\pm0.40$   & $0.505\pm0.092^{d}$ 
     & $0.75\pm0.03$   & $10.5\pm0.6$     
     & $4.09\pm0.06$ 
     & $67^{+26}_{-22}$ & 8\\
\hline
\end{tabular} 
\end{center}
*) our adopted values. Our $H_0$ is estimated
for the $\Omega_0=1$ and $\lambda_0=0$ model.
1) Yamashita  (1994),
2) Birkinshaw et al. (1981), 
3) Carlstrom, Joy, \& Grego (1996),
4) Rephaeli (1995),
5) Elbaz, Arnaud \& B\"{o}hringer (1995),
6) Arnaud et al. (1992), Wilbanks et al. (1994), 
7) Birkinshaw \& Hughes (1994),
8) Myers et al. (1996).

a) the observed radio frequency $\widetilde\nu=28.7$GHz,
b) $\widetilde\nu=136$GHz, 
c) $\widetilde\nu=20.3$GHz, 
d) $\widetilde\nu=32$GHz, 
e) in this cluster we use $S_\X(0)= (5.2 \pm
0.83)\times10^{-6}$erg$\,$cm$^{-2}\,$s$^{-1}\,$ arcmin$^{-2}$
directly in eq.(9) to compute $H_0$ and $d_\A$,
f) estimated assuming $\Omega_0=1$ and $\lambda_0=0$.
\end{table}
}

\setcounter{figure}{0}

\vspace*{-1cm}
\begin{figure}
\begin{center}
   \leavevmode\psfig{figure=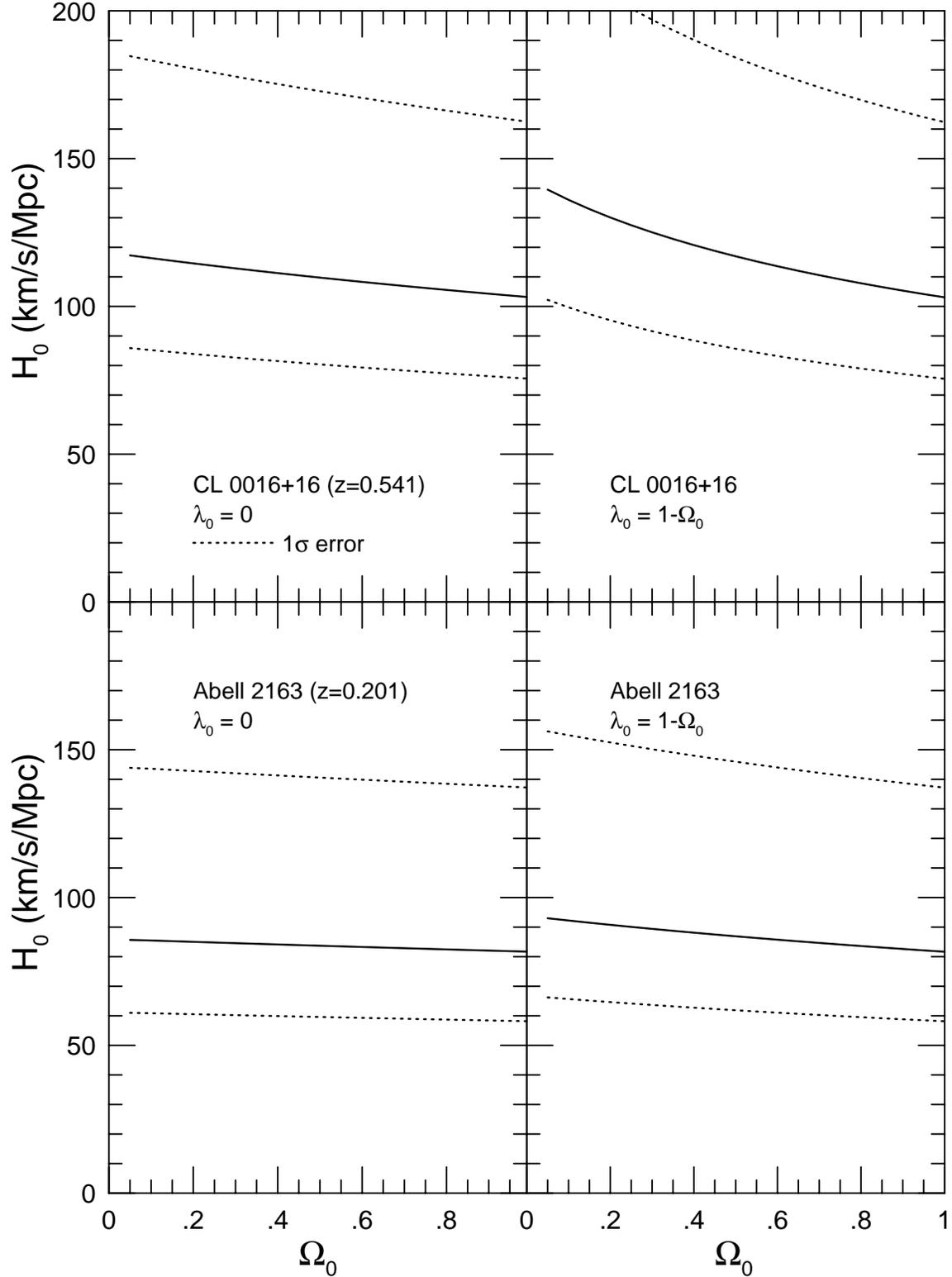,width=15cm}
\end{center}
\vspace*{-1cm}
\caption{
  $H_0$ estimated from the SZ effect as a function of $\Omega_0$ in
  the case of $\lambda_0=0$ (left panels) and $\lambda_0=1-\Omega_0$
  (right panels) assuming the parameters in Table 1. Upper panels are
  for CL0016+16, while lower panels for A2163.  The solid and dotted
  lines represent the central value and $\pm 1\sigma$ error ranges,
  respectively.
\label{fig:h0omega0}
}\end{figure}

\vspace*{-3cm}
\begin{figure}
\begin{center}
   \leavevmode\psfig{figure=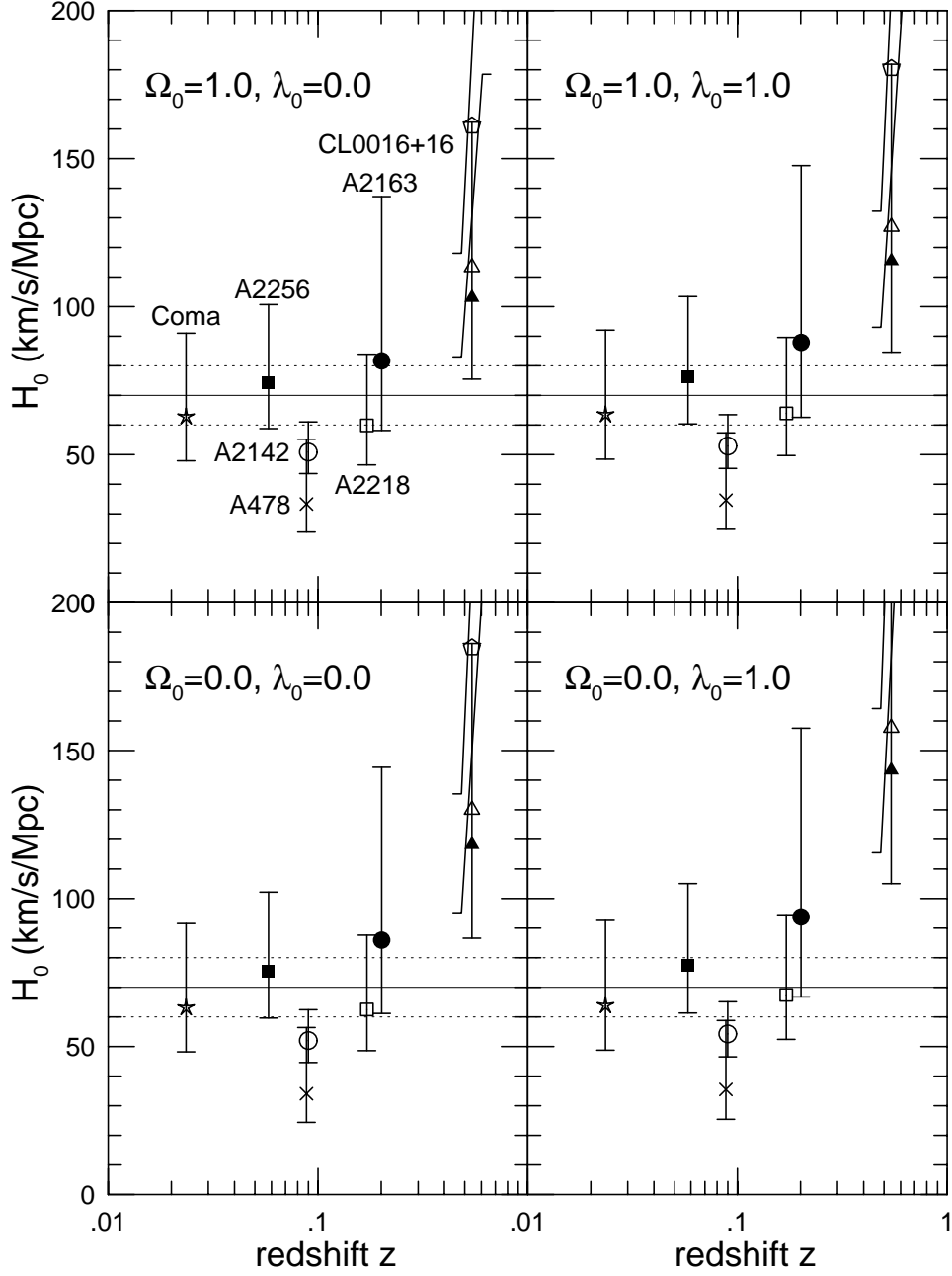,width=16cm}
\end{center}
\vspace*{-2cm}
\caption{
  $H_0$ estimated from the SZ effect for seven clusters with $1\sigma$
  error bars in models with ($\Omega_0$, $\lambda_0$)= (1.0, 0.0),
  (1.0, 1.0), (0.0, 0.0) and (0.0, 1.0).  For CL0016+16, we plot
  values with different $\beta$ and $\theta_c$: $\beta_\X=\beta_r =
  0.5, \theta_\X=\theta_r= 0.5'$ (open pentagon), $\beta_\X=0.5,
  \beta_r = 2/3, \theta_\X=0.5', \theta_r= 0.7'$ (open triangle), and
  $\beta_\X=\beta_r = 2/3, \theta_\X=\theta_r= 0.7'$ (filled
  triangle). The error bars for the latter two are artficially tilted
  to avoid confusion.
\label{fig:h0z}
}\end{figure}

\begin{figure}
\begin{center}
   \leavevmode\psfig{figure=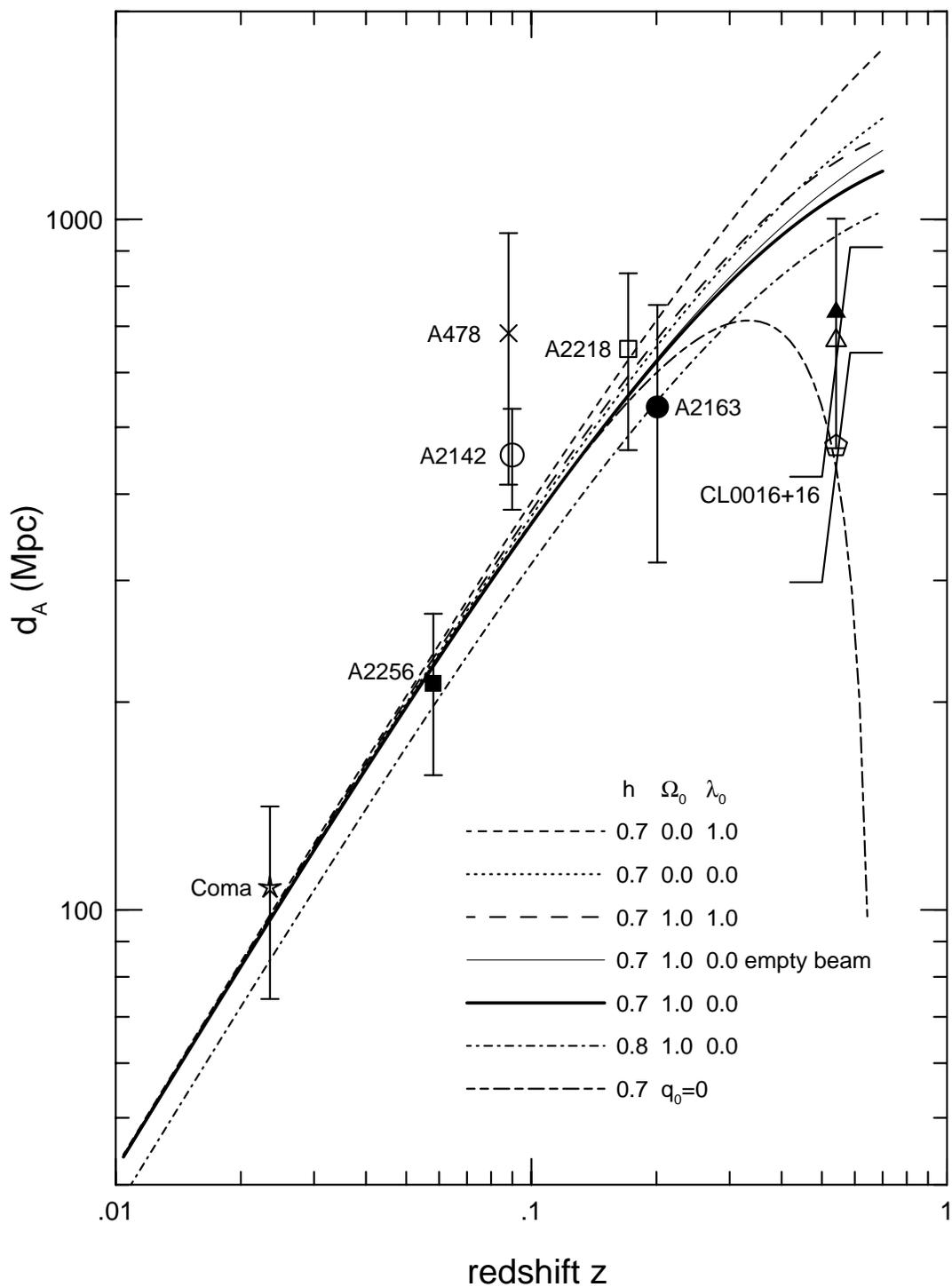,width=14cm}
\end{center}
\caption{
The angular diameter distance $d_\A$ estimated from the SZ
  effect as a function of $z$ for seven clusters (symbols).
 Lines indicate theoretical curves for several sets of cosmological
 parameters ($h$, $\Omega_0$, $\lambda_0$). The meanings of the symbols 
 are the same as Fig. 2.
\label{fig:daz}
}\end{figure}

\end{document}